# Fermi liquid behavior and colossal magnetoresistance in layered MoOCl$_2$

Zhi Wang,[1,#] Meng Huang,[1,#] Jianzhou Zhao,[2,#] Cong Chen,[2] Haoliang Huang,[3] Xiangqi Wang,[4] Ping Liu,[1] Jianlin Wang,[3] Junxiang Xiang,[1] Chao Feng,[1] Zengming Zhang,[5] Xudong Cui,[6] Yalin Lu,[1] Shengyuan A. Yang,[2] and Bin Xiang[1,*]

[1]*Hefei National Research Center for Physical Sciences at the Microscale, Department of Materials Science & Engineering, CAS Key Lab of Materials for Energy Conversion, University of Science and Technology of China, Hefei, Anhui 230026, China*
[2]*Research Laboratory for Quantum Materials, Singapore University of Technology and Design, Singapore 487372, Singapore*
[3]*National Synchrotron Radiation Laboratory, CAS Center for Excellence in Nanoscience, University of Science and Technology of China, Hefei, Anhui 230026, China*
[4]*Department of Physics, University of Science and Technology of China, Hefei, Anhui 230026, China*
[5]*The Centre for Physical Experiments, University of Science and Technology of China, Hefei, Anhui 230026, China*
[6]*Sichuan New Materials Research Center, Institute of Chemical Materials, CAEP, Chengdu, Sichuan, 610200 China*

[#]These authors contributed to the work equally
[*]Corresponding author: binxiang@ustc.edu.cn

A characteristic of a Fermi liquid is the $T^2$ dependence of its resistivity, sometimes referred to as the Baber law. However, for most metals, this behavior is only restricted to very low temperatures, usually below 20 K. Here, we experimentally demonstrate that for the single-crystal van der Waals layered material MoOCl$_2$, the Baber law holds in a wide temperature range up to ~120 K, indicating that the electron-electron scattering plays a dominant role in this material. Combining with the specific heat measurement, we find that the modified Kadowaki-Woods ratio of the material agrees well with many other strongly correlated metals. Furthermore, in the magneto-transport measurement, a colossal magneto-resistance is observed, which reaches ~350% at 9 T and displays no sign of saturation. With the help of first-principles calculations, we attribute this behavior to the presence of open orbits on the Fermi surface. We also suggest that the dominance of electron-electron scattering is related to an incipient charge density wave state of the material. Our results establish MoOCl$_2$ as a strongly correlated metal and shed light on the underlying physical mechanism, which may open a new path for exploring the effects of electron-electron interaction in van der Waals layered structures.

**INTRODUCTION**

Electron-electron interaction is fundamental to condensed matter physics and underlies many fascinating physical effects. In the most striking cases, it can open energy gaps in an original metallic spectrum and lead to exotic phases such as superconductivity [1,2] and charge density wave state [3,4]. Meanwhile, even if the system remains in a metallic state, the electron-electron interaction often significantly modifies its various thermodynamic and transport properties. For example, in Landau's Fermi liquid theory, the electron-electron interaction gives rise to a contribution $\sim T^2$ [5-7]. This important result is sometimes referred to as the Baber law [8]. However, in most real metallic materials, this characteristic $T^2$ dependence can only be observed at very low temperature, usually below 20 K, because the electron-phonon scattering can easily dominate the electron-electron scattering at elevated temperatures and the former features the familiar Bloch $T^5$ law or the linear $T$ dependence [9]. Nevertheless, in titanium dichalcogenides, particularly $TiS_2$ and $TiTe_2$, it was reported in experiment that the $T^2$ dependence could hold in a much wider range. For $TiTe_2$, the range extends to 60 K [10], while for $TiS_2$, the upper bound can even reach 400 K [11]. This unusual behavior has not been fully understood yet. And it is natural to ask whether this kind of behavior can be found in other material families beyond the titanium dichalcogenides.

Molybdenum oxide dichloride ($MoOCl_2$) is a typical van der Waals (vdW) layered material. It belongs to a big family of layered oxide dichlorides $MOCl_2$ (M=V, Ta, Nb, Mo, Os) [12]. Interestingly, $MoOCl_2$ (and other members of this family) was synthesized more than 50 years ago and demonstrated with a metallic conductivity down to 4.2 K. However, its underlying physical mechanism of electrical conduction and magnetoresistance have not been carefully investigated yet [13,14]. In this work, we discover that MoOCl$_2$ is a new example which features strong electron-electron (*e-e*) scattering. Through transport measurement, we show that the material exhibits the $T^2$ scaling of the resistivity in a wide range from 3 K to 123 K. Above ~133 K, the temperature scaling crosses over to a linear $T$ dependence due to electron-phonon (*e*-ph) scattering, without going through the usual $T^5$ scaling window. Combined with the specific heat measurement, we find that the material's modified Kadowaki-Woods ratio is consistent with many other strongly correlated metals, such as the transition metals and the heavy fermion metals. Furthermore, in magneto-transport measurement, we find a colossal non-saturating magneto-resistance (MR) which reaches 350% at 9 T without going through any phase transition. We attribute this behavior to the presence of open orbits on the material's Fermi surface, which is confirmed by the first-principles calculation. Our result reveals a new platform for the studying the fascinating physics of electron-electron scattering, and shed light on the possible mechanisms behind the exotic behavior.

**RESULTS AND DISCUSSION**

MoOCl$_2$ has a layered structure with monoclinic symmetry (space group *C2/m*). Each unit layer consists of three atomic layers: the central Mo-O atomic layer is sandwiched by two layers of Cl atoms, as shown in Fig. 1(a). In 1963, Schäfer *et al*. firstly reported the synthesis of MoOCl$_2$ crystals by chemical transport, confirming that MoOCl$_2$ crystallizes in the NbOCl$_2$-type structure [13]. In this work, our single-crystal samples are synthesized by a vapor transport method (see Materials and Methods for the details). The transmission electron microscopy (TEM) image and X-ray diffraction (XRD) pattern (Fig. S1 of the Supplemental Material [15]) reveal the high quality of our samples. Then, a dry transfer method is applied to prepare the MoOCl$_2$ based electrical device for the four-probe transport measurement [Fig. 1(b)].

In Fig. 1(c), we plot the temperature dependence of the resistivity measured in as-prepared MoOCl$_2$. The resistivity increases with temperature, clearly indicating a metallic character. The fitting analysis in Fig. 1(c) reveals that the $T^2$ dependence in the resistivity can reach a remarkably high temperature $T_0 \sim 123$ K (light pink region), and above ~133 K, the scaling directly crosses over to the linear $T$ dependence (light blue region). The narrow region (light cyan) between the $T^2$ and $T$ represents the crossover region. This result indicates that the electron-electron scattering plays a dominant role over a wide temperature range for transport in MoOCl$_2$. Moreover, there is an unusual downturn of resistivity that deviated from the linear near 300 K, which may due to the inevitably temperature hysteresis between the rotator sample mounts and system chamber during the initial cooling stage in practical measurements.

Below the cross-over temperature $T_0$, the resistivity varies as $\rho(T) = \rho_0 + AT^2$, where $\rho_0$ is the residual resistivity, and the coefficient $A$ encodes the electron-electron scattering effects and is typically proportional to $1/\varepsilon_F^2$ [16]. To obtain a more accurate number of $A$, in Fig. 2(a), we plot the resistivity as a function of $T^2$, such that $A$ corresponds to the slope of the curve (square symbol). The obtained result is $A = 9.7$ n$\Omega$ cm K$^{-2}$. In Fig. 2(b), we also show the measurement results when a magnetic field is applied (in the perpendicular direction of the device layer). Throughout our experiment, we do not observe any signal of structural phase transition induced by the temperature variation or the applied fields. One observes that as the magnetic field increases from 0 to 9 T, the shape of the $\rho$ vs $T^2$ curve is more or less unchanged. Fitting shows that the value of $A$ slightly decreases from 9.7 (at 0 T) to 8.3 n$\Omega$ cm K$^{-2}$ (at 9 T). Meanwhile, the cross over temperature increases from 123 K (at 0 T) to 155 K (at 9 T) (The results are robust and easily reproducible based on more devices in Fig. S3). This trend is somewhat expected, because the magnetic field suppresses the ZXT kinetic energy of the electrons hence enhances the electron-electron interaction effects [17].

Another characteristic behavior of Fermi liquid is the linear $T$ dependence of the specific heat at low temperatures. The material specific heat can typically be fitted with the relation $c = \gamma T + \beta T^3$, where the first term comes from the electronic contribution, and the second term is from the phonon contribution [18]. (There could exist an additional thermal correction $\propto T^3 \ln T$ [19-21]. However, our current data does not give a conclusive evidence for this correction term (Fig. S4).) In Fig. 2(c), we plot the measured specific heat for MoOCl$_2$, which can be well fitted by this formula. The $\gamma$ and $\beta$ extracted in the low temperature range take the values of 13.57 mJ/mol K$^2$ and 3.1 mJ/mol K$^4$, respectively. In the simplest model, $\gamma$ is connected to the effective mass of the electrons via the relation $\gamma = m^* n^{1/3} (k_B/\hbar)^2 (\pi/3)^{2/3}$, from which we can obtain that $m^* \approx 6.16\ m_e$. Meanwhile, $\beta$ is connected to the Debye temperature $\Theta_D$, as in the Debye model, the phonon contribution to the specific heat is given by $c_{\text{ph}} = \left(\frac{12\pi^4}{5}\right) N k_B \left(\frac{T}{\Theta_D}\right)^3$ at low temperature [22]. The Debye temperature estimated for MoOCl$_2$ is about 95 K.

Now let us come back to Fig. 1(c). We have seen that the in MoOCl$_2$, the temperature scaling for the resistivity directly crosses over from $T^2$ to $T$. This is in contrast to the conventional behavior, e.g. in alkali metals, where a $T^5$ scaling region is observed in between [23]. The above estimated Debye temperature helps us to understand this exotic feature. The electron-electron scattering dominates the transport below the cross-over temperature $T_0$. In MoOCl$_2$, $T_0$ is very high and even higher than $\Theta_D$. Because the Bloch $T^5$ scaling only appears for $T$ lower than $\Theta_D$, this $T^5$ window is in fact buried in the range below $T_0$, and is overwhelmed by the $T^2$ contribution from the electron-electron scattering [24].

The coefficient $A$ in the scaling relation for resistivity and the coefficient $\gamma$ in the scaling relation for specific heat both contain electron-electron interaction effects. Their ratio $\alpha = A/\gamma^2$, known as the Kadowaki-Woods ratio, is an important parameter which probes the relationship between the electron-electron scattering rate and the renormalization of the effective mass. It usually takes the similar value among a similar class of materials. For example, for many transition metals including Fe and Ni, $\alpha_{\text{TM}} \approx 0.4$ µΩ cm mol² K²/J²; while for many heavy fermion compounds, $\alpha_{\text{HF}} \approx 10$ µΩ cm mol² K²/J². Here, according to our experimental result, we find that $\alpha_{\text{MoOCl}_2} \approx 52.7$ µΩ cm mol² K²/J² for MoOCl$_2$, which is more than five times larger than the typical value for heavy fermion compounds. A comparison of the Kadowaki-Woods ratios (on a log-log scale) is reproduced for a number of materials (Fig. S2). The large value of $\alpha$ (larger than for $\alpha_{\text{HF}}$ for heavy fermion compounds) is conventionally interpreted as an indicator of a strongly correlated metal [25]. This is consistent with our observation of the unusually high cross-over temperature $T_0$.

There has been effort to unify the different values of Kadowaki-Woods ratio for different classes of materials. Noticeably, Jacko *et al.* have proposed a modified form of Kadowaki-Woods ratio which takes into account the carrier density ($n$) and the effective dimensionality ($d$) of the system [26]. The modified definition reads $\tilde{\alpha} = (A/\gamma^2)f_d(n)$, where $f_d(n)$ is an extra correction factor. Because MoOCl₂ is a vdW layered material, we take its effective dimension to be two. For transport in an in-plane direction, we have $f_2(n) = \frac{n^2}{\pi\hbar^2 l}$, where $l$ is the interlayer spacing. For MoOCl₂, we have $n = 2.8 \times 10^{22} cm^{-3}$, and $l \sim 6$ Å, so the modified $\tilde{\alpha} = 3.88 \times 10^{26}$ Ω K² s⁶ g⁻⁴ Å⁻⁸. In Fig. 2(d), we present a plot of the modified ratio among the different materials. Indeed, one observes that in the $A$ vs $\gamma^2/f_d$ plot, all the materials nicely fall onto the same straight line, and the result of MoOCl₂ is similar to other strongly correlated metals.

In addition, we have also investigated the magnetic properties and the MR of MoOCl₂ (Fig. S5). Here, MR is defined as MR= 100% × $[R(H) - R(0)]/R(0)$, where $R(H)$ and $R(0)$ represent the resistances with and without magnetic field, respectively. In Fig. 3(a), we plot the MR measured at 3 K as a function of the applied magnetic field for two field directions. When the field is in-plane and parallel to the current flow, the MR is very small (~5% at 9 T). Remarkably, when the field is out-of-plane (i.e. along the crystal c-axis), one observes a colossal positive MR of 350%, which is one order of magnitude higher than other reported Fermi liquid oxides [27]. Furthermore, the MR exhibits no sign of saturation in the applied field range. In Fig. 3(b) and 3(c), we plot the MR as a function of the angle between the applied magnetic field and the out-of-plane direction (c-axis), which can be well fitted by a function of $|\cos(\theta)|$. This clearly shows that only the out-of-plane component of the field is effective in generating the MR, which reflects the character of vdW layered structure of MoOCl₂.

What is the possible mechanism behind this colossal MR? The contribution from the quantum interference effect is unlikely to play an important role here because its typical size is orders of magnitude smaller than what we observed. Some other mechanisms, associated with carrier spin polarization [28], or extremely high mobility [29,30], have been proposed to explain large MR. In our system, the carrier density determined by Hall measurement is $\sim 10^{22}$ cm$^{-3}$ and the average mobility is $\sim 50$ cm$^2$/Vs (Fig. S6). So these mechanisms are also not relevant here. One notes that the MR has a strong variation when the direction of $B$ field is rotated. Combined with its non-saturating behavior, it hints at the possible existence of open orbits on the Fermi surface. This speculation is confirmed by our first-principles calculations. In Fig. 4, we plot the band structure and the Fermi surface obtained from the calculation. One observes that the low-energy bands in MoOCl$_2$ are dominated by the Mo 4$d$ orbitals. The band dispersion is small along the out-of-plane direction, consistent with the vdW layered structure. Importantly, there exist several pieces of Fermi surface, which traverse the Brillouin zone [Fig. 4(d)]. These pieces accommodate open orbits lying inside the layer plane. For example, in Fig. 4(e), we show the Fermi surface in the $k_z = 0$ plane, in which one can clearly observe the open orbits along the $k_x$ (i.e., [100]) direction. The presence of these open orbits explains the huge non-saturating MR observed in experiment.

Additionally, we mention that there exist van Hove singularities slightly above the Fermi level, associated with the saddle points at Z and Y points in the band structure (see the Supplemental Material for more details, Fig. S7). These points correspond to the $n = 2$ van Hove singularity [31,32], and give a logarithmic contribution to the density of states (DOS). This enhancement in DOS should further promote the electron-electron interaction effects.

We have a few remarks before closing. First, our experiment and analysis have demonstrated that the electron-electron scattering dominates over an unusually wide temperature range for transport in MoOCl$_2$. This is connected with the strong electronic correlation in transition metal $d$ orbitals. Meanwhile, it may also benefit from the ineffectiveness of the electron-phonon scattering in affecting the electron transport. In electron-phonon scattering, the effective electron momentum relaxation requires that $k_D < 2k_F$ (where $k_D$ is the Debye wave vector). In MoOCl$_2$, we estimate that $k_D \approx 2.2k_F$ (Supplemental Material Sec. A [33-35]), hence the phonon resistivity is somewhat suppressed [11]. This naturally favors the dominance of electron-electron scattering.

Second, the crystal structure of MoOCl$_2$ is similar to the previously studied titanium dichalcogenides. They are all layered materials, with each unit layer consisting of three atomic layers. This shed light on the possible reasons behind their unusual transport properties. In titanium dichalcogenides, there is experimental evidence of charge-density-wave (CDW) phases under certain conditions [36]. In the MoOCl$_2$ investigated here, we do not observe signature of a metal-insulator transitions. The material remains metallic for all the field and temperature ranges studied. Nevertheless, it is possible that the material is close to a CDW transition, or CDW has gapped part of (not the whole) Fermi surface. Indeed, from the band structure result in Fig. 4(a), one clearly observes that the dispersion of the low-energy bands is strong only in the $k_y$ direction, i.e., parallel to the Mo-O chain in real space (Fig. S8); whereas the dispersion along other directions are relatively flat. This reflects a strong anisotropy of the system: in terms of these low-energy states, it resembles a quasi-1D system consisting of weakly coupled Mo-O chains, which hence is prone to CDW formation. This argument is also evidenced by the several almost parallel Fermi surfaces observed in Fig. 4(d), which exhibit the nesting feature. As a result, the electron-electron interaction can be enhanced by the strong fluctuations into the incipient CDW state. This interesting possibility deserves more future studies.

**SUMMARY AND CONCLUSION**

In summary, we have experimentally demonstrated unusual transport properties in the vdW layered material of MoOCl$_2$. We find that the Baber law can hold to a temperature beyond 120 K for MoOCl$_2$, much higher than most metallic materials. And the scaling directly crosses over into the linear $T$ dependence, without going through a region of the Bloch $T^5$ scaling. The modified Kadowaki-Woods ratio of MoOCl$_2$ is in good agreement with the values of many strongly correlated metals. In addition, we find a colossal non-saturating MR in MoOCl$_2$, reaching 350% at 9 T, which can be attributed to the open orbits on the Fermi surface. Our results reveal MoOCl$_2$ as a promising platform to explore the physics of electron-electron interactions. Given its vdW layered structure, the material may be thinned down to the 2D form and combined with other layered materials to form vdW heterostructures. This may open a path to further probe the interplay between electron-electron interaction and the many fascinating properties of 2D materials in future studies.

**Materials and Methods**

**Sample synthesis**

The MoOCl$_2$ single crystals were prepared by a low-temperature chemical vapor transport (CVT) method. The stoichiometric high-purity MoO$_3$ powder (99.95%) and α-MoCl$_3$ single-crystal flakes were weighted into evacuated quartz tubes (ϕ8 mm×12 cm, $10^{-2}$ Pa). The sealed ampoules were placed into a tube furnace with a ~100 K thermal gradient from the center to the edge, and subsequently kept at 573 K for 20 days. Large amounts of ribbon-like crystals can be obtained in the low temperature zone with maximum dimension size of 2 mm×0.05 mm×0.02 mm. MoOCl$_2$ ribbon-like crystals are uniformly golden yellow and radiated from the core with a sea-urchin-like morphology.

**Device fabrication**

Both four-probe and Hall bar devices of layered MoOCl$_2$ were fabricated directly by an all-dry transfer process method. The MoOCl$_2$ flake were mechanically exfoliated on the surface of the viscoelastic stamp (polydimethylsiloxane, PDMS) and identified by optical microscopy. Then the stamp was pressed against the acceptor surface of SiO$_2$/Si substrate with pre-patterned Ti/Au electrodes and peeled off very slowly. Subsequently, the selected MoOCl$_2$ flake was successfully transferred onto the target pre-patterned electrodes. Finally, multilayer *h*-BN was used as protective encapsulation for MoOCl$_2$ sample.

**Measurements**

All of the transport and specific heat measurements were performed in the Physical Property Measurement System (PPMS) manufactured by Quantum Design (QD) with a base temperature of 1.8 K and a magnetic field of up to 9 T. The angular dependence of magnetoresistance was measured using the PPMS rotator option. The magnetic properties were measured using a superconducting quantum interferometric device (SQUID, Quantum Design).

**First-principles calculations**

The first-principle calculations were carried out based on the density-functional theory (DFT) as implemented in the Vienna *ab initio* Simulation Package (VASP), by using the Projector Augmented Wave (PAW) method. The Generalized Gradient Approximation (GGA) with Perdew-Burke-Ernzerhof (PBE) realization was adopted for the exchange-correlation potential. 6 electrons ($4d^55s$) in Mo, 6 electrons ($2s^22p^4$) in O, and 7electrons ($3s^23p^5$) in Cl were treated as valence electrons. The plane-wave cutoff energy was set to 500 eV. The Γ centered k-point mesh of size 15 x 15 x 15 was used for the Brillouin Zone (BZ) sampling. The spin-orbit coupling effect was included in our calculations. The experimental lattice parameters with *a*=12.721 Å, *b*=3.755 Å, *c*=6.524 Å and *ß*=104.86º were used in our DFT calculation. The energy convergence criterion was set to be $10^{-7}$ eV.

**Acknowledgments**

This work was supported by the joint fund of the National Natural Science Foundation Committee of the China Academy of Engineering Physics (NSAF) (U1630108), the National Key Research and Development Program of China (2017YFA0402902), and the Singapore Ministry of Education AcRF Tier 2 (MOE2017-T2-2-108). This research was partially carried out at the USTC Center for Micro and Nanoscale Research and Fabrication.

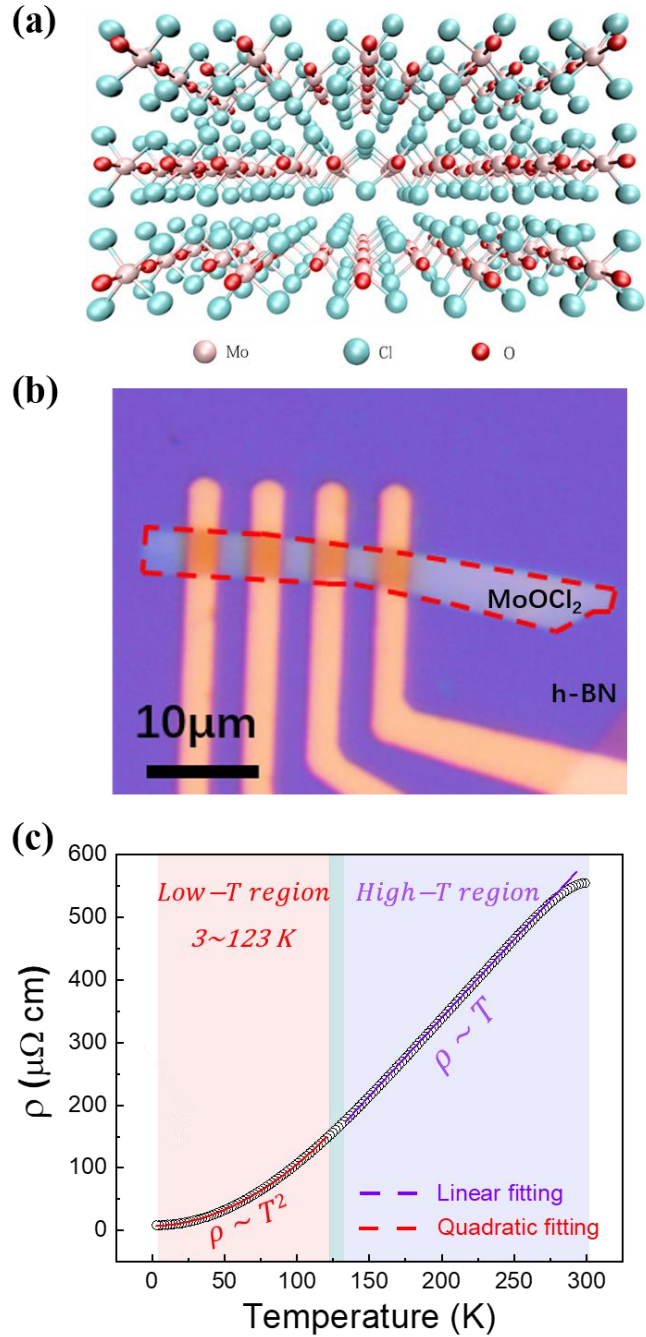

FIG. 1. Structure and transport of MoOCl$_2$. (a) Schematic perspective view of the layered crystal structure of MoOCl$_2$, with Mo-O-Mo chains parallel to the b-axis. (b) Optical image of a four-probe device based on MoOCl$_2$ flake (thickness: ~20 nm) with *h*-BN as protective encapsulation. (c) MoOCl$_2$ temperature-dependent resistivity plot and the corresponding fitting curves with no magnetic field applied. The crossover temperature is 123 K.

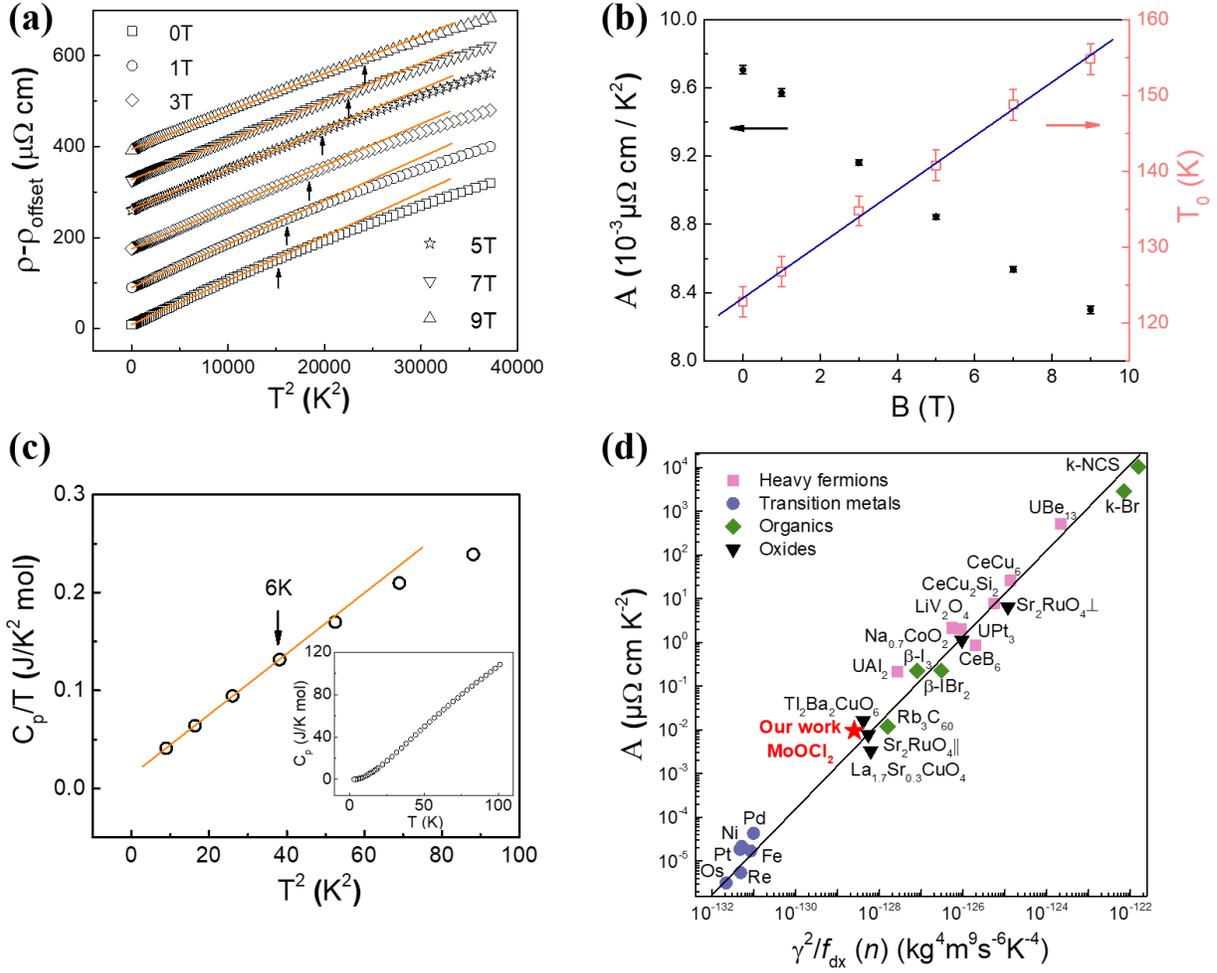

FIG. 2. 2D Fermi-Liquid behavior in layered MoOCl$_2$. (a) Temperature dependence of the in-plane electrical resistivity at different magnetic fields plotted as ρ-ρ$_{offset}$ vs T$^2$, where ρ$_{offset}$ is a constant arbitrary offset chosen for clarity of display. The solid lines represent the quadratic fit with ρ=ρ$_0$+AT$^2$ below a crossover temperature T$_0$ marked by arrows. (b) Magnetic field dependence of the *A*-coefficients and the upper temperature limit (T$_0$) in (a) with solid fitting line, respectively. (c) Plot of $C_p/T$ vs $T^2$ with a linear fitting at a low temperature of 6 K by $C_p/T=\beta T^2+\gamma$ in solid line. The temperature dependence of the MoOCl$_2$ specific heat from 3 to 100 K in the inset. (d) The modified Kadowaki-Woods ratio plots of our MoOCl$_2$ (red star) and other strong electron-electron correlated materials.

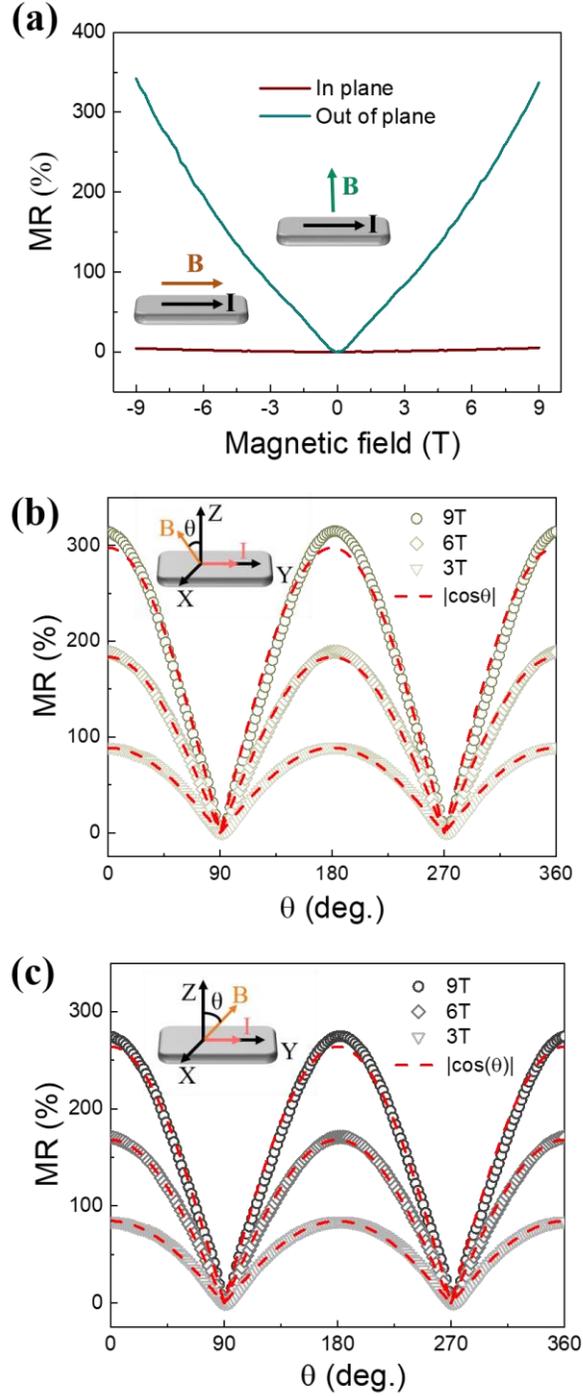

FIG. 3. Colossal positive magnetoresistance and magnetic anisotropy of MoOCl$_2$. (a) Longitudinal magnetoresistance at 3 K with a magnetic field parallel (wine) and perpendicular (dark cyan) to the a-b plane of MoOCl$_2$. (b) and (c) The angular dependence of magnetoresistance at 3K under the magnetic field of 3T, 6T and 9T. The red dash lines represent the fitting curves by a function of $|\cos(\theta)|$. Insets: the relationship between the applied current and magnetic field in the measurements.

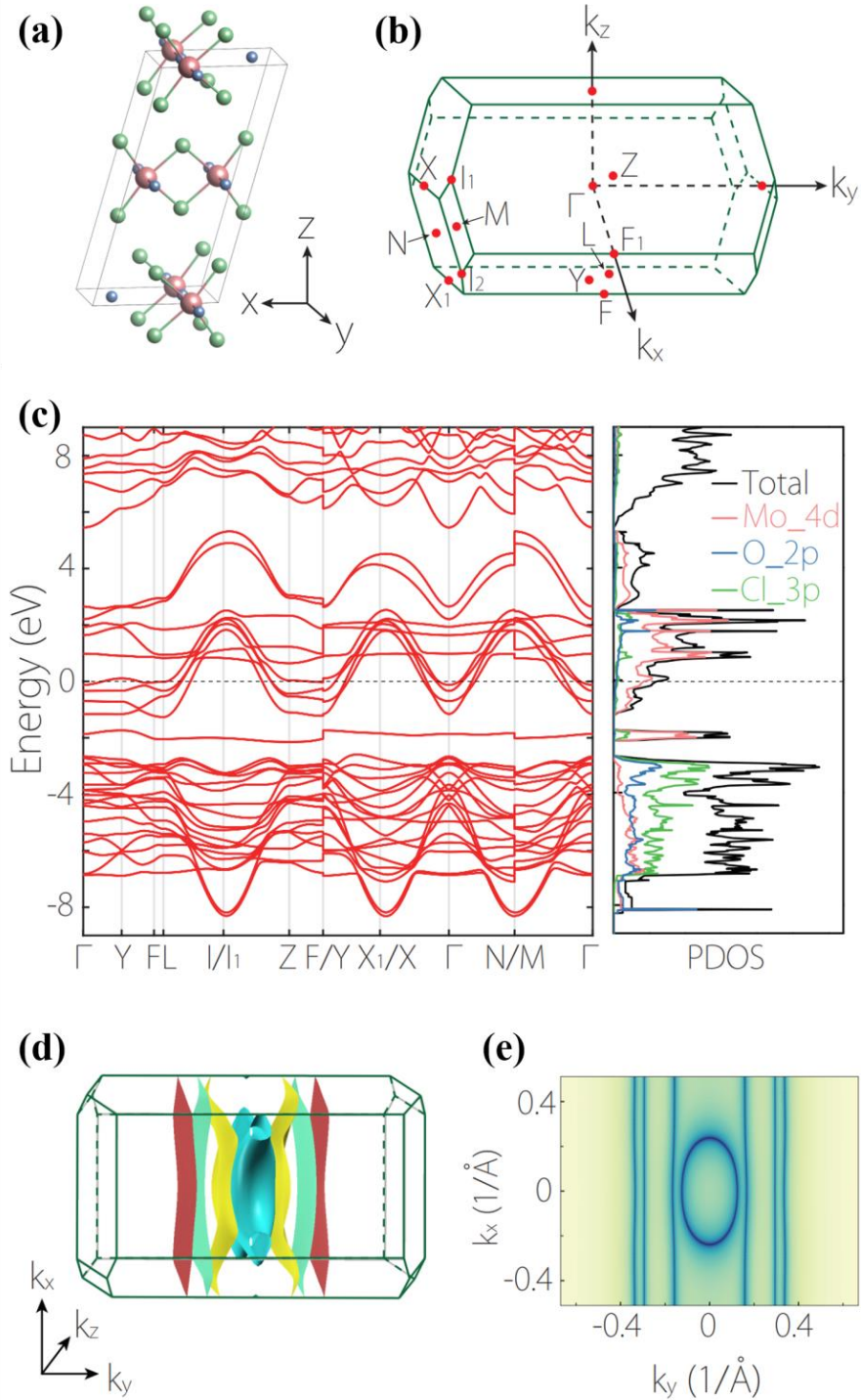

FIG. 4. Calculated electronic structure of MoOCl$_2$. (a) Layer crystal structure of MoOCl$_2$. (b) Reciprocal unit cell with the high-symmetry points. (c) Band structure of MoOCl$_2$ along high symmetry lines obtained by first-principles calculations. The right panel shows the orbital projected density of states. (d) Fermi surfaces in the first BZ. The four bands crossing the Fermi level are illustrated with four different colors. (e) Fermi surface contours in the $k_z$=0 plane.